# Effect of doping on performance of organic solar cells

V. A. Trukhanov, V.V. Bruevich, D.Yu. Paraschuk

*International Laser Center and Faculty of Physics,*

*M.V. Lomonosov Moscow State University, Moscow 119991, Russia*

(Dated: December 1, 2011)

Conventional models of planar and bulk heterojunction organic solar cells have been extended by introducing doping in the active layer. We have studied the performance of organic solar cells as a function of dopant concentration. For bulk heterojunction cells, the modeling shows that for the most studied material pair (poly-3-hexylthiophene, P3HT, and phenyl-$C_{61}$-butyric acid methyl ester, PCBM) doping decreases the short-circuit current density ($J_{SC}$), fill factor (*FF*) and efficiency. However, if bulk heterojunction cells are not optimized, namely, at low charge carrier mobilities, unbalanced mobilities or non-ohmic contacts, the efficiency can be increased by doping. For planar heterojunction cells, the modeling shows that if the acceptor layer is *n* doped, and the donor layer is *p* doped, the open-circuit voltage, $J_{SC}$, *FF* and hence the efficiency can be increased by doping. Inversely, when the acceptor is *p* doped, and the donor is *n* doped; *FF* decreases rapidly with increasing dopant concentrations so that the current-voltage curve becomes *S* shaped. We also show that the detrimental effect of nonohmic contacts on the performance of the planar heterojunction cell can be strongly weakened by doping.

## I. INTRODUCTION

The efficiency of organic solar cells has been increasing rapidly for recent years. Meanwhile, the key electronic processes in them, e.g., charge generation, recombination, and transport, are still under intense discussion. Device modeling is an efficient tool to understand and optimize the performance of organic solar cells. Hence, adequate device models are needed. The first models for current-voltage characteristics of organic solar cells were based on the standard approaches that justified themselves for inorganic solar cells.[1,2] However, these models do not take into account the mechanism of charge generation and recombination in organic semiconductors and are not applicable to describe the observed features in current-voltage characteristics of organic solar cells.[3] In fact, the photophysics in organic solar cells is essentially different from that in inorganic ones. In organic semiconductors, light absorption results in excitons with the binding energy much higher than the thermal energy. To generate the photocurrent, the excitons need to be dissociated into free charge carriers; this can be done by using a heterojunction of type II. The heterojunction can be formed at the contact of two organic semiconductors with different electron affinities and ionization potentials – donor and acceptor.

The simplest case is a planar heterojunction formed at the interface between the donor and acceptor layers. In bilayer organic solar cells (i.e., with planar heterojunction), only those photons contribute to the photocurrent that are absorbed within the exciton diffusion length from the interface. Barker *et al.*[4] have proposed a numerical model of bilayer organic solar cells. This model takes into account drift and diffusion of charge carriers, the effect of space charge on the electric field in the device, and generation/recombination of free charges. The generation and recombination in the model occur through bound electron-hole pairs at the donor-acceptor interface with an electric field-dependent dissociation rate. The model reproduces many important features of the measured current-voltage characteristics of polyfluorene-based bilayer organic photovoltaic devices.

On the other hand, it is widely accepted that bilayer organic solar cells have limited efficiency because the exciton diffusion length is significantly less than the optical absorption length (~100 nm). Most efficient organic solar cells are based on bulk heterojunction[5] that implies very high interface area distributed over the heterojunction volume. The bulk heterojunction can be realized in donor-acceptor blends with separated donor and acceptor phases so that the characteristic phase separation length would be of the order of the exciton diffusion length.[6,7] Therefore, the majority of photogenerated excitons can dissociate into free charges and contribute to the photocurrent. To model bulk heterojunction organic solar cells, it was proposed to start with the metal-insulator-metal picture,[8] where the bulk heterojunction layer is considered as one virtual semiconductor with the properties of both the donor and acceptor. Based on this approach, Koster *et al.*[9] have proposed a numerical device model that consistently describes the current-voltage characteristics of bulk heterojunction cells. Later this model was extended by taking into account injection barriers[10] and reduced surface recombination velocities[11] at the contacts with electrodes.

The models of organic solar cells do not take into account doping of active layers as they presume that organic semiconductors are intrinsic, i.e., undoped. However, organic semiconductors are known to be not pure: they have defects and impurities, with some of them being charged. The charged defects can act as dopants and affect the exciton dissociation and charge transport in the active layer of organic solar cells.[12,13] The measured conductivities in organic semiconductors are vastly higher than expected for intrinsic defect-free semiconductors.[14-20] This suggests that the organic semiconductors can be unintentionally doped.[13] It is naturally to suggest that both intended and unintentional doping could strongly affect the key processes in organic solar cells.

In this paper, the effect of doping on the performance of planar and bulk heterojunction organic solar cells is studied by numerical modeling. We extend the models of bulk[9,10] and planar[4] heterojunction organic solar cells by introducing doping of the photoactive layer(s). We take into account doping of the active layer(s) and study how the dopants influence the energy diagrams, electric field distribution, and current density-voltage (*J-V*) characteristics of bilayer and bulk heterojunction solar cells. We assume that doping does not influence the exciton dynamics; specifically, the dopants do not quench the excitons. In first approximation, this assumption implies that the exciton diffusion length, $L_{ex}$, should be less than the average distance between the charged dopants, $N^{-1/3}$. Accordingly, for typical $L_{ex} \sim 10$ nm, $N$ should be less than $10^{24}$ m$^{-3}$. Note that the photoinduced free carrier density generated in a 100-nm-thick active layer under one-sun illumination typically reaches $10^{22}$ m$^{-3}$ (e.g., see Fig. 11 below). Moreover, the experimental data on bulk[21] and planar[22] heterojunction solar cells under concentrated sunlight show that the short-circuit current is linear at least up to 10 suns. Therefore, exciton quenching by free carriers can be neglected at least for a doping density of $10^{23}$ m$^{-3}$. As will be shown below, doping considerably affects the performance of organic solar cells at the doping density in the range $10^{22}$–$10^{24}$ m$^{-3}$. As a starting material system we use the most studied polymer-fullerene pair: poly(3-hexylthiophene) (P3HT) and phenyl-C$_{61}$-butiryc acid methyl ester (PCBM). For optimized bulk heterojunction cells, we find that doping decreases the efficiency. However, for nonoptimized cells the efficiency can be increased by doping. For bilayer solar cells, we show that if the layers are doped by majority carriers, the photocurrent increases as a result of increasing interface electric field. Nevertheless, if the layers are doped by minority carriers, the fill factor decreases, and the *J-V* characteristics become *S*-shaped. Our model also demonstrates that the negative effect of non-ohmic contacts on the performance of bilayer solar cells can be partially compensated by majority carrier doping.

## II. MODEL

### A. Doping

Organic semiconductors can have highly polarizable defects and impurities. However, their majority may be not ionized because of low dielectric constant.[13] In this study, we neglect the influence of the non-ionized defects and impurities because they are neutralized and do not create a macroscopic electric field. The ionized defects and impurities result in free charge carriers, i.e., they act as dopants.

There are two types of dopants: *n* and *p* type. Energy levels of *n*-type dopants are in the semiconductor band gap near the lower edge of the conduction band $E_c$, and they easily give electrons to the conduction band, but the dopants itself are charged positively. Analogously, *p*-type dopants have energy levels near the upper edge of the valence band $E_v$, and they easily accept electrons from the valence band, i.e., generate holes. The Poisson equation for the electric field strength *E* in the presence of dopants is

$$\frac{dE}{dx} = \frac{e}{\varepsilon \varepsilon_0}(p - n + N_n - N_p), \tag{1}$$

where $N_n$ and $N_p$ are *n*- and *p*-type concentrations of ionized dopants, respectively, *e* is the electron charge, *p* and *n* are the concentrations of free holes and electrons, correspondingly, $\varepsilon$ is the dielectric constant, $\varepsilon_0$ is the vacuum permittivity, and *x* is the spatial coordinate. Generally, the both types of dopants are contained in organic semiconductors. But inevitably, one type dominates: free electrons from *n*-type dopants recombine with holes from *p*-type dopants leaving the one type of charge carriers and a lot of compensated dopants.

The concentration of uncompensated dopants can be estimated from the conductivity $\sigma = e\mu n$. The charge carrier mobility $\mu$ in organic semiconductors is typically in the range $10^{-9} - 10^{-6}$ m$^2$/(V s). The measured conductivity is usually in the range of $10^{-10} - 10^{-5}$ S/m for thin films of molecular organic semiconductors[14-16] and $10^{-6} - 10^{-3}$ S/m for conjugated polymers.[17-20] These values are greatly higher than expected for intrinsic pure semiconductors suggesting that the conductivity can be determined by doping, and the concentration of ionized uncompensated dopants can reach $10^{24}$ m$^{-3}$.

### B. Bulk heterojunction solar cells

For modeling the effect of doping on bulk heterojunction cells we start from the model introduced by Koster *et al.*[9] This model is based on the metal-insulator-metal picture: the active layer with bulk heterojunction is considered as an effective semiconductor that has properties of both the donor and acceptor materials. This semiconductor has $E_v$ corresponding to the highest occupied molecular orbital (HOMO) of the donor and $E_c$ corresponding to the lowest unoccupied molecular orbital (LUMO) of the acceptor.

The numerical model of bulk heterojunction cell is based on equations:

$$-\frac{d^2\varphi}{dx^2} = \frac{dE}{dx} = \frac{e}{\varepsilon\varepsilon_0}(p - n + N_n - N_p), \qquad (2)$$

$$\frac{1}{e}\frac{dj_n}{dx} = P(E)G - (1-P(E))\alpha(np - n_{int}^2), \qquad (3a)$$

$$-\frac{1}{e}\frac{dj_p}{dx} = P(E)G - (1-P(E))\alpha(np - n_{int}^2), \qquad (3b)$$

$$j_n = en\mu_n E + \mu_n kT\frac{dn}{dx}, \qquad (4a)$$

$$j_p = ep\mu_p E - \mu_p kT\frac{dp}{dx}. \qquad (4b)$$

The unknown functions are the concentrations of free electrons $n(x)$ and holes $p(x)$, and the electric potential $\varphi(x)$. The Poisson equation [Eq. (2)] describes the dependence of the electric potential and field on space charge, with doping being taken into account. The charge generation and recombination processes are described by the current continuity equations for electrons and holes [Eqs. (3a) and (3b)]. $G$ is the generation rate of bound electron-hole pairs, $P(E)$ is the probability of bound electron-hole pair dissociation, and $n_{int}$ is the intrinsic concentration of charge carriers, $n_{int}^2 = N_c N_v \exp(-E_g/kT)$. $N_{c,v}$ are the effective densities of states in conduction and valence bands, and $E_g$ is the effective band gap that is the energy difference between the acceptor LUMO and the donor HOMO. Recombination is assumed to be bimolecular, the recombination constant $\alpha$ is given by Langevin: [23,24]

$$\alpha = \frac{e}{\varepsilon\varepsilon_0}(\mu_n + \mu_p), \qquad (5)$$

where $\mu_n$ and $\mu_p$ are the electron and hole mobilities, correspondingly. The current densities of electrons $j_n$ and holes $j_p$ are presented as a sum of drift and diffusion current densities in Eqs. (4a) and (4b). Diffusion is assumed to obey the Einstein relation,[25] and hence the diffusivity is proportional to the temperature $kT$, $k=1.38\times10^{-23}$ J/K is the Boltzmann constant. The total current density through the active layer is a sum of the electron and hole current densities $J = j_n + j_p$.

The charge generation in this model is a three-step process. First, the absorbed light generates excitons. Second, they diffuse to the donor-acceptor interface, where charge separation takes place, and as a result bound electron-hole pairs are formed at the interface. Third, these pairs either recombine monomolecularly with the decay rate $k_f$ or dissociate on free charges with rate $k_{diss}$. The latter rate depends on the electron-hole distance $a$ and the electric field strength $E$. This dependence was derived by Braun,[26] who based it on Onsager's theory for field-dependent dissociation rate constants in weak electrolytes:[27]

$$k_{diss}(a,E) = \frac{3\alpha}{4\pi a^3} e^{-\frac{e^2}{4\pi\varepsilon\varepsilon_0 akT}} J_1(2\sqrt{-2b})/\sqrt{-2b}, \qquad (6)$$

where $\alpha$ is determined by Eq. (5), $b = e^3|E|/(8\pi\varepsilon\varepsilon_0 k^2 T^2)$, and $J_1$ is the Bessel function of first order. The probability of dissociation of bound electron-hole pairs is given by

$$p(a,E) = \frac{k_{diss}(a,E)}{k_{diss}(a,E) + k_f}. \qquad (7)$$

As the bulk heterojunction is a disordered system, it is assumed that the electron-hole pair distance $a$ is not constant throughout the active layer.[28] As a result, Eq. (7) should be integrated over a distribution of electron-hole pair distances:

$$P(E) = \int_0^\infty p(a,E) f(a) da, \qquad (8)$$

where $f(a)$ is assumed to be a normalized distribution function given by[29]

$$f(a) = \frac{4}{\sqrt{\pi} a_0^3} a^2 e^{-a^2/a_0^2}, \qquad (9)$$

where $a_0$ is the average electron-hole pair distance.

The boundary conditions at the contacts $x=0$ and $x=L$ ($L$ is the active layer thickness) for unknown functions are

$$\varphi(0) = 0, \quad \varphi(L) = \Phi_2 - \Phi_1 + V, \qquad (10a)$$

$$n(0) = N_c \exp\left(\frac{\chi - \Phi_1}{kT}\right), \quad n(L) = N_c \exp\left(\frac{\chi - \Phi_2}{kT}\right), \qquad (10b)$$

$$p(0) = N_v \exp\left(\frac{\Phi_1 - \chi - E_g}{kT}\right), \quad p(L) = N_v \exp\left(\frac{\Phi_2 - \chi - E_g}{kT}\right). \qquad (10c)$$

$\Phi_1$ and $\Phi_2$ are the work functions of electrodes, $V$ is the voltage applied to the electrodes, $\chi$ is the electron affinity. The resulting boundary value problem can be solved numerically using the method proposed by Gummel.[30,31]

To obtain the *J-V* characteristics, one needs to solve Eqs. (2)-(4) using boundary conditions [Eq. (10)] with various values of *V*.

### C. Bilayer organic solar cells

To model the effect of doping on bilayer solar cells we modify the model proposed by Barker et al.[4] Our model is based on the equations similar to those for the bulk heterojunction cells:

$$\begin{cases} -\dfrac{d^2\phi_i}{dx^2} = \dfrac{dE_i}{dx} = \dfrac{e}{\varepsilon_i\varepsilon_0}(p_i - n_i + N_{n,i} - N_{p,i}) \\ \dfrac{1}{e}\dfrac{dj_{n,i}}{dx} = -\alpha_i\left(n_i p_i - n_{int}^2\right) \\ \dfrac{1}{e}\dfrac{dj_{p,i}}{dx} = \alpha_i\left(n_i p_i - n_{int}^2\right) \\ j_{n,i} = en_i\mu_{n,i}E_i + \mu_{n,i}kT\dfrac{dn_i}{dx} \\ j_{p,i} = ep_i\mu_{p,i}E_i - \mu_{p,i}kT\dfrac{dp_i}{dx} \end{cases} \quad (11)$$

where index $i=1,2$ corresponds to one of the layers.

The boundary conditions at the electrodes are the same to those for the bulk heterojunction model [Eq. (10)]. Also it is necessary to set matching conditions for the unknown functions at the donor-acceptor interface. The generation of free charges at the interface is taken into account in the matching conditions for the electron and hole current densities, $j_n$ and $j_p$:

$$j_{n,2} - j_{n,1} + ePG_X - \dfrac{e^2 h(1-P)(\mu_{n,1}+\mu_{p,2})}{3\varepsilon\varepsilon_0}n_1 p_2 - \dfrac{e^2 h(1-P)(\mu_{n,2}+\mu_{p,1})}{3\varepsilon\varepsilon_0}n_2 p_1 = 0, \quad (12a)$$

$$j_{p,1} - j_{p,2} + ePG_X - \dfrac{e^2 h(1-P)(\mu_{n,1}+\mu_{p,2})}{3\varepsilon\varepsilon_0}n_1 p_2 - \dfrac{e^2 h(1-P)(\mu_{n,2}+\mu_{p,1})}{3\varepsilon\varepsilon_0}n_2 p_1 = 0, \quad (12b)$$

where $G_X$ is the surface generation rate of bound electron-hole pairs, $P$ is determined by Eq. (7), $h$ is the effective separation distance between the layers. The last two terms in Eqs. (12a) and (12b) describe recombination of free electrons and holes through the interface.

The dependence of the dissociation rate of bound electron-hole pairs on the electric field follows from treatment of Jonscher[33] and is given by expression

$$k_{diss}(E) = \begin{cases} k_{diss}(0)\dfrac{2}{M}\left[e^M\left(1-\dfrac{1}{M}\right)+\dfrac{1}{M}\right], & \text{if } E|_{interface} < 0 \\ k_{diss}(0)\dfrac{4}{M^2}\left(1-e^{-\frac{M^2}{4}}\right), & \text{if } E|_{interface} > 0, \end{cases} \quad (13)$$

where

$$M = \frac{e}{kT}\sqrt{\frac{e|E|}{\pi\varepsilon\varepsilon_0}}\bigg|_{\text{interface}} \tag{14}$$

and $k_{diss}(0)$ is the dissociation rate at zero electric field.

If for bulk heterojunctions $k_{diss}$ depends on the absolute value of the electric field, for planar heterojunctions the sign of the interface electric field is important. If the electric field is collinear to the dipole moment of a bound electron-hole pair, the dissociation rate will increase with increasing $E$. But if the dipole moment and the electric field are opposite, the dissociation rate will be suppressed.

Because the dissociation and recombination rates of bound electron-hole pairs enter in Eq. (7) as a ratio, the parameter of the model will be the ratio of the zero-field dissociation rate to the decay rate

$$K = \frac{k_{diss}(0)}{k_f}. \tag{15}$$

### III. DOPED BULK HETEROJUNCTION SOLAR CELLS

#### A. Optimized cell

First, we present the results for optimized bulk heterojunction solar cell based on the most studied material pair: P3HT and PCBM with their weight ratio 1:1. We term optimized cells those with the parameters – charge carrier mobilities, electrode work functions, etc. – chosen so that the efficiency is maximal. The input parameters used in modeling bulk heterojunction cells are listed in Table I. The generation rate of bound electron-hole pairs $G$ corresponds to an incident light intensity of 1000 W/m$^2$ with spectrum AM1.5. Below we discuss the results for different values of dopant concentrations.

TABLE I. The input parameters used in modeling bulk heterojunction solar cells.

| Parameter | Symbol | Numerical value |
| --- | --- | --- |
| Active layer thickness | $L$ | 100 nm |
| Temperature | $T$ | 300 K |
| Effective band gap | $E_g$ | 1.05 eV |
| Electron affinity | $\chi$ | 4 eV |
| Left electrode work function | $\Phi_1$ | 5.05 eV |
| Right electrode work function | $\Phi_2$ | 4 eV |
| Dielectric constant | $\varepsilon$ | 4 |
| Electron and hole mobilities | $\mu_{n,p}$ | $10^{-7}$ m$^2$/(V s) |
| Effective density of states | $N_{c,v}$ | $10^{26}$ m$^{-3}$ |
| Generation rate | $G$ | $9 \times 10^{27}$ m$^{-3}$ s$^{-1}$ |
| Electron-hole pair distance | $a_0$ | 1.3 nm |
| Decay rate | $k_f$ | $10^4$ s$^{-1}$ |

Figure 1 shows the *J-V* characteristics under illumination calculated at different levels of *p* doping. For *n* doping, the results are similar. One can see that doping results in substantial decreasing of the short-circuit current density $J_{SC}$. Fill factor *FF* decreases as well although the open-circuit voltage $V_{OC}$ slightly increases. As a result, the efficiency decreases with doping.

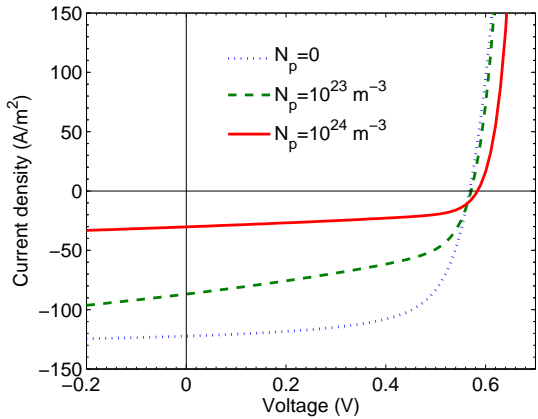

FIG. 1. *J-V* characteristics at different *p*-type dopant concentrations $N_p$.

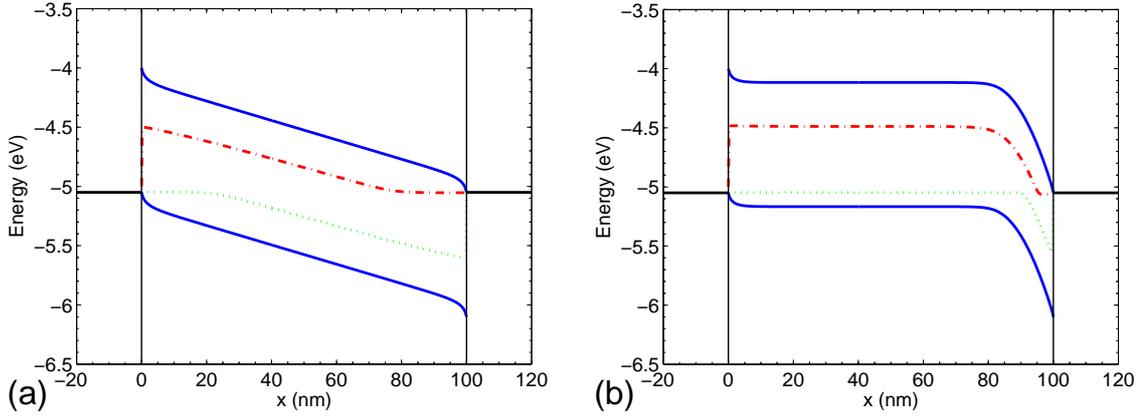

FIG. 2. Energy band diagrams under short-circuit conditions for undoped active layer (a) and *p* doped one at $N_p=10^{24}$ m$^{-3}$ (b). The vertical black lines denote the active layer-electrode interfaces, the solid lines are $E_c$ and $E_v$, the dash-dotted/dotted line is the electron/hole quasi-Fermi level. The Fermi levels of electrodes are denoted by horizontal black solid lines.

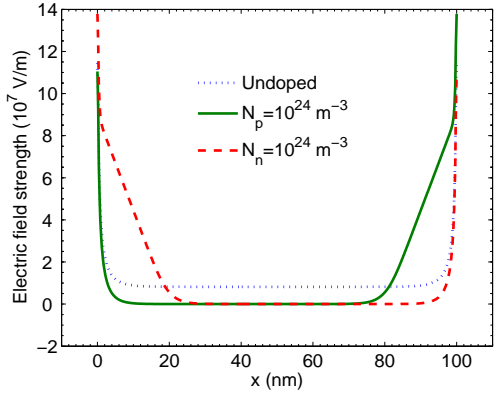

FIG. 3. Absolute value of the electric field strength in the active layer under short-circuit conditions for undoped, *p*-doped, and *n*-doped bulk heterojunction cell.

To explain the doping effect, consider the cell in short circuit. Figure 2 compares the calculated energy band diagrams for undoped and *p*-doped active layers. The charge density in the undoped active layer is approximately zero, so the energy levels have an almost constant slope [see Fig. 2(a)], and a non-zero electric field is present throughout the layer. Figure 3 shows the distribution of electric field strength in the undoped and doped active layers. At *p* doping, there is abundance of free holes in the active layer. Because of the strong difference between the work functions of the electrode and the semiconductor at the right contact, holes go from the active layer to the electrode. This leads to formation of negative space charge in the active layer near the right contact (cathode). This space charge creates a high electric field near the right contact (see Fig. 3, solid line), so the energy levels are bended in this part of the active layer [see

Fig. 2(b)], i.e., a Schottky barrier is formed. In the residue of the active layer (at $x<80$ nm), due to high concentration of free charges, the electric field is almost zero, and the energy levels are horizontal. For *n*-doped bulk heterojunction, the effect of doping is similar, but the electric field is concentrated near the left electrode (anode) (Fig. 3, dashed line).

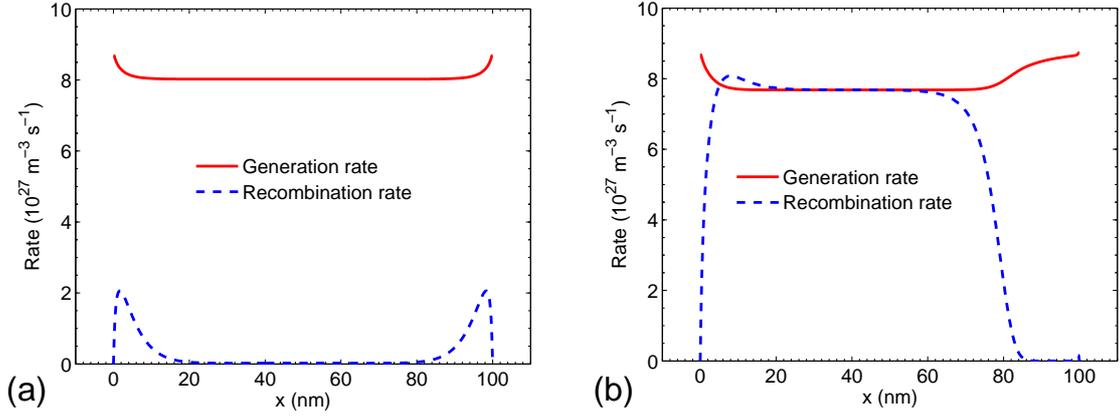

FIG. 4. Generation and recombination rates of free charges in short-circuit conditions in the undoped active layer (a) and *p*-doped one at $N_p=10^{24}$ m$^{-3}$ (b).

Figure 4 displays the generation and recombination rates of free charges in the undoped and *p*-doped active layer. The generation rate of free charges in the undoped bulk heterojunction cell is constant throughout the active layer as a result of the constant electric field, and the recombination rate is negligible because the majority of generated free charges quickly escapes the active layer driven by this field. In the doped active layer, generation of free charges occurs effectively only in a small region of the active layer where the electric field is not zero [near the right electrode in Fig. 4(b)]. In the residue of the active layer, where the electric field is very low, generation of charges is compensated by their recombination.

Figure 5 shows the dependences of $J_{SC}$, *FF*, $V_{OC}$, and efficiency on dopant concentration. At low dopant concentration ($<10^{22}$ m$^{-3}$), the solar cell parameters almost do not depend on doping. However, at higher doping, the efficiency decreases and drops by about four times at $10^{24}$ m$^{-3}$. This decrease is mainly due to reduction of $J_{SC}$ while the influence of doping on $V_{OC}$ is weak as in open-circuit the electric field is much lower than in short-circuit. Therefore the electric field and the processes of charge generation, transport, and recombination are almost unaffected by doping. *FF* varies with doping nonmonotonically: it decreases by ~20% at weak doping ($<10^{23}$ m$^{-3}$) and then increases by ~10% at higher doping. We assign the lowering of *FF* to decreasing of the ratio between charge generation and recombination rates in the major part of the active layer. On the other hand, this ratio increases near the right electrode [see Fig. 4(b)] so that charges generated there give more photocurrent. As these charges do not recombine, they can result in a *J-V* curve with higher *FF* at high doping.

The calculations were also performed for *n* doping. The dependencies of $J_{SC}$, *FF*, $V_{OC}$ and efficiency are the same as long as the input parameters are symmetric with respect to electrons and holes (i.e., the same mobilities $\mu_n=\mu_p$, densities of states $N_c=N_v$, injection barriers at the contacts, etc.). We report the results for asymmetric input parameters below.

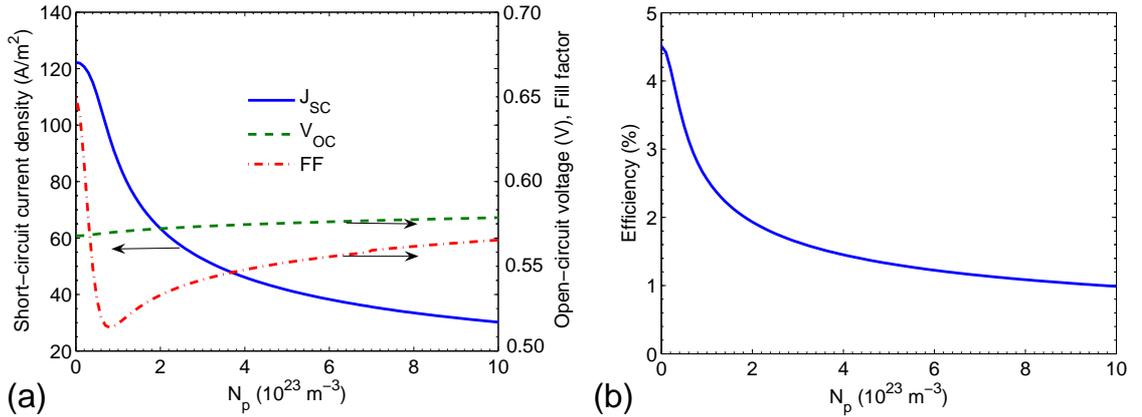

FIG. 5. Short-circuit current density $J_{SC}$, open-circuit voltage $V_{OC}$, fill factor *FF* (a) and efficiency (b) vs *p*-type dopant concentration $N_p$.

In summary, both *p* and *n* doping of optimized bulk heterojunction polymer-fullerene solar cells deteriorates their performance, as their $J_{SC}$ and *FF* decrease. However, if the parameters of bulk heterojunction are not optimal, doping can enhance the efficiency. Specifically, there are cases of low charge carrier mobilities, unbalanced electron and hole mobilities, and non-ohmic contacts that are considered below.

## B. Low mobilities

Low charge carrier mobilities are known to limit the efficiency of organic solar cells. Influence of charge mobility on the performance of bulk heterojunction cells was studied earlier using a similar numerical model.[10] Here, we consider low charge carrier mobilities $\mu_n=\mu_p=10^{-11}$ m$^2$/(V s), that are lower by four orders of magnitude than in the optimal bulk heterojunction (the previous section), and study the effect of doping on the solar cell performance. The other parameters are taken from Table I.

Figure 6 compares the *J-V* characteristics at low mobilities calculated for different *p*-doping levels. Doping results in increasing $J_{SC}$, $V_{OC}$, *FF* and hence the efficiency. To explain this behavior, consider the processes of charge generation and recombination under short-circuit conditions.

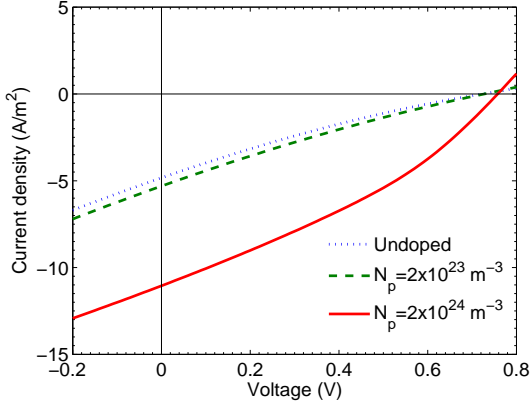

FIG. 6. Current voltage characteristics at low charge mobilities [$\mu_n=\mu_p=10^{-11}$ m$^2$/(V s)] for undoped and p-doped bulk heterojunction.

The electric field distribution in the doped/undoped active layer is similar to that shown in Fig. 3 for the doped/undoped optimized cell. Low charge mobilities result in a low bound electron-hole pair dissociation rate, $k_{diss}(E)$, according to Eqs. (5) and (6). This decreases the dissociation probability $P$ [Eq. (8)] in the electric field of undoped cell resulting in low photocurrent. But, with doping, the electric field increases near one of the electrodes (as a result of Schottky barrier) increasing $P$ as well. Therefore, the generation rate of free charges increases, and their recombination rate decreases near the right electrode of $p$-doped cell [Fig. 7(b)]. In the undoped cell, generation of free charges is almost compensated by their recombination [Fig. 7(a)] because the charges are very slow, and so only their minority reaches the electrodes. For the $p$-doped cell, the charge generation rate near the right electrode ($90<x<100$ nm) is several times higher as compared with the undoped cell, and the recombination is suppressed [Fig. 7(b)].

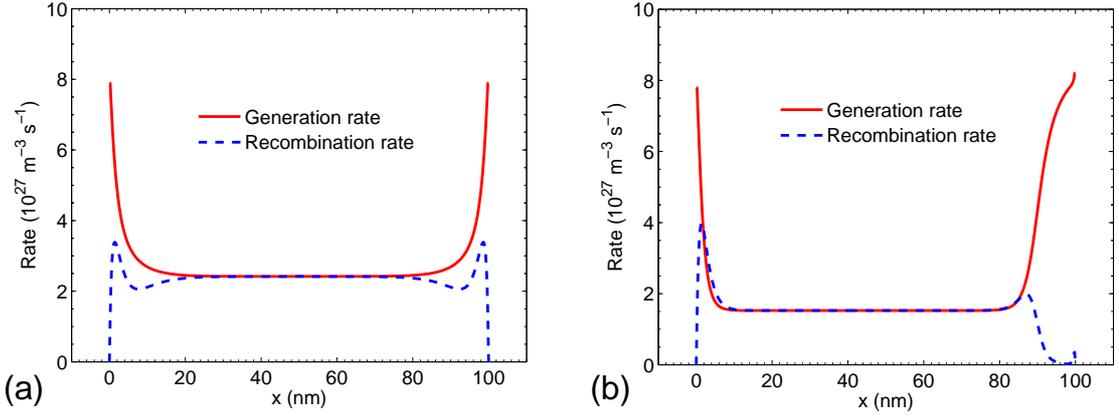

FIG. 7. Free charge generation and recombination rates at $\mu_n=\mu_p=10^{-11}$ m$^2$/(V s) under short-circuit conditions for undoped active layer (a) and p-doped one at $N_p=2\times10^{24}$ m$^{-3}$ (b).

Thus, for low charge carrier mobilities, the efficiency of bulk heterojunction cells increases at doping. However, the maximum efficiency is not very high as compared with that of the undoped optimized cell. Figure 8 shows $J_{SC}$, $V_{OC}$, $FF$, and efficiency as functions of $p$-dopant concentration. The efficiency of the undoped cell is more than one order of magnitude less than of the optimized cell in accordance with the earlier data.[10] With doping, $J_{SC}$ peaks at $N_p=2\times10^{24}$ m$^{-3}$, but the efficiency reaches its maximum at somewhat higher doping ($N_p=3.4\times10^{24}$ m$^{-3}$) as $FF$ monotonically increases with doping. As a result, the efficiency increases by four times at doping. It should be noted that with $n$-doping the results are the same.

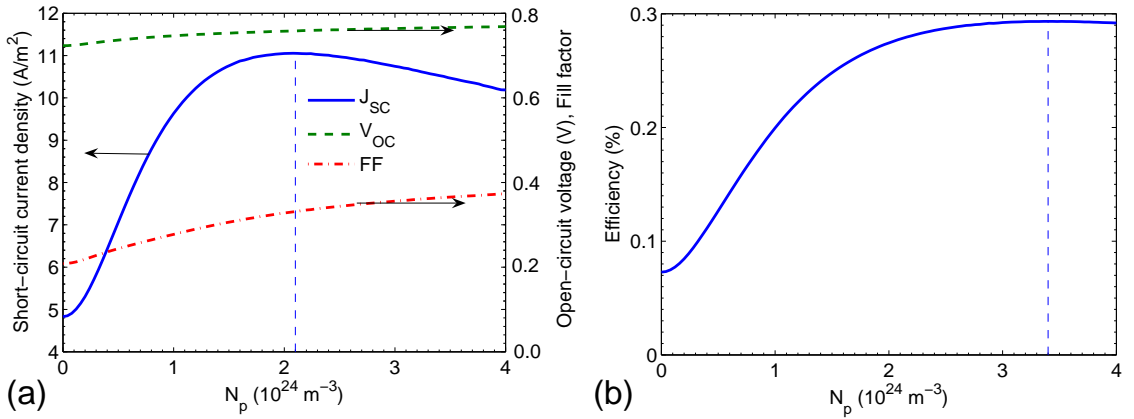

FIG. 8. Short-circuit current density $J_{SC}$, open circuit voltage $V_{OC}$, fill factor $FF$ (a) and efficiency (b) vs $p$-dopant concentration $N_p$ at low charge carrier mobilities $\mu_n=\mu_p=10^{-11}$ m$^2$/(V s). Vertical dashed lines denote the maximal $J_{SC}$ and efficiency.

### C. Unbalanced mobilities

In this section, we show that the photocurrent and efficiency of bulk heterojunction solar cells can be increased by doping if the electron and hole mobilities are unbalanced. Unbalanced mobilities are typical for organic solar cells, and their difference can be several orders of magnitude. As a result, the photocurrent decreases due to a space-charge effect of slower charge carriers.[33,34]

Figure 9 shows the calculated *J-V* characteristics of bulk heterojunction cell with balanced (optimized cell) and unbalanced ($\mu_p/\mu_n=10^{-2}$) mobilities. $J_{SC}$ is maximal at the balanced mobilities and decreases by 25% at the unbalanced ones. However, in the *p*-doped cell, $J_{SC}$ increases by 12% at $N_p=6.1\times10^{22}$ m$^{-3}$. Thus, doping by slower carriers can partly weaken the negative effect of unbalanced charge mobilities. Figure 10 plots the dependences of the main parameters of bulk heterojunction cell with unbalanced mobilities on the dopant concentration of both types. Now in contrast to the balanced mobilities [Fig. 10(b), black dashed line] these dependences are not symmetrical with respect to the type of doping. For unbalanced mobilities, $J_{SC}$ is maximal at $N_p=5.6\times10^{22}$ m$^{-3}$, but the maximum efficiency is observed at a higher concentration ($N_p=6.1\times10^{22}$ m$^{-3}$), as *FF* slightly increases with *p* doping.

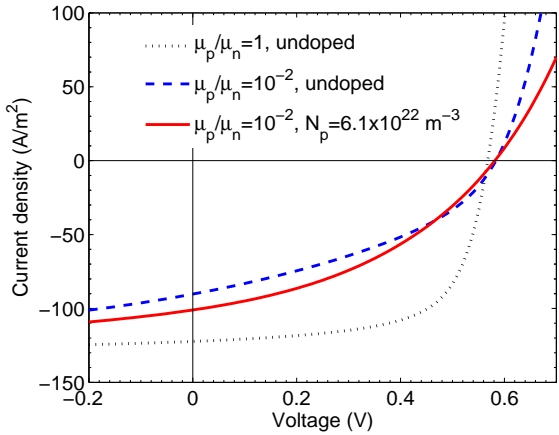

FIG. 9. *J-V* characteristics at unbalanced mobilities of electrons and holes [$\mu_p/\mu_n=10^{-2}$, $\mu_n=10^{-7}$ m$^2$/(V s)] for undoped (dashed line) and p-doped (red solid line) bulk heterojunction. The black dotted line denotes the *J-V* characteristic for undoped cell with balanced mobilities ($\mu_p=\mu_n$).

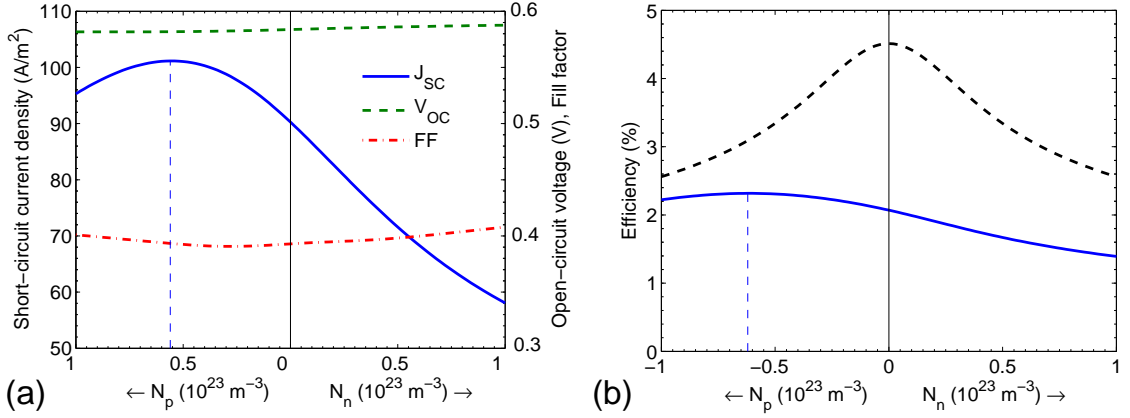

FIG. 10. Short-circuit current density $J_{SC}$, open-circuit voltage $V_{OC}$, fill factor $FF$ (a) and efficiency (b) vs dopant concentration of $n$-type ($N_n$) and $p$-type ($N_p$) at unbalanced mobilities [$\mu_p/\mu_n=10^{-2}$, $\mu_n=10^{-7}$ m$^2$/(V s)]. The black dashed line in panel (b) denotes the efficiency at balanced mobilities ($\mu_p=\mu_n$). The vertical dashed lines indicate the maximal $J_{SC}$ and efficiency.

Increasing the photocurrent with doping for a cell with unbalanced carrier mobilities can be explained as follows. Because of mobility imbalance, the generated free charges leave the active layer with different rates, so the slower charge carriers are accumulated in it. Figure 11 depicts the concentrations of charges and their recombination rate in the undoped and $p$-doped active layer at a mobility ratio of $\mu_p/\mu_n=10^{-2}$. In the bulk of active layer, the hole concentration is about two orders of magnitude higher than the electron one. As a result, space charge is formed in the active layer. This space charge creates an additional electric field, which reduces the electric field formed by the electrodes, thereby enhancing recombination of free charges (Fig. 11, dotted line) and reducing the photocurrent. If doping is introduced, $p$-type dopants can partly compensate the space charge in the active layer [see Eq. (1)], thus suppressing the recombination of free charges (Fig 11, dash-dotted line). Therefore, $J_{SC}$ and the efficiency can be increased by doping.

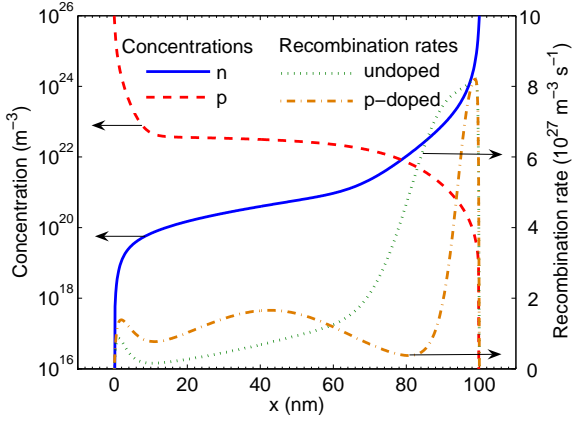

FIG. 11. Concentrations of electrons $n$ and holes $p$ in the active layer at unbalanced mobilities ($\mu_p/\mu_n =10^{-2}$) at short-current conditions; their recombination rate for undoped active layer and p-doped one at $N_p=6.1\times10^{22}$ m$^{-3}$.

### D. Non-ohmic contacts

Non-ohmic contacts are formed if the electrode Fermi levels differ strongly from $E_v$ (the donor HOMO) at the anode contact or $E_c$ (the acceptor LUMO) at the cathode contact so that the injection barriers are formed. This is known to limit the cell performance.[35,36] Here we consider the effect of doping for a bulk heterojunction cell with non-ohmic anode contact. As a transparent anode, indium tin oxide (ITO) with work function within the range 4.2 – 4.75 eV[37] is usually used. Nevertheless, the polymers most studied in organic photovoltaics have the HOMO energy around 5 eV, and semiconducting polymers with higher HOMO are in great demand. To align the anode work function with the donor HOMO (e.g., P3HT), a thin film of appropriate material with high work function is usually placed between the ITO and active layers, e.g., poly(3,4-ethylenedioxithiophene):ploy(styrenesulfonate) (PEDOT:PSS). Therefore, if the ITO is used without the additional layer, the anode contact can be nonohmic. To study the effect of doping on bulk heterojunction solar cells with nonohmic contacts, we have chosen the anode (hole-injecting electrode) work function $\Phi_1$=4.3 eV as compared with $\Phi_1$=5.05 eV for the optimized cell (see Table I). As a result, an injection barrier for holes is formed at the left contact, so the contact is nonohmic.

Figure 12 compares the $J$-$V$ characteristics calculated for undoped and $p$-doped bulk heterojunction solar cells with the nonohmic contact. For comparison, the $J$-$V$ curve of the optimized solar cell is also shown. As the difference of electrode work functions is lower than that for ohmic contacts, the electric field in the active layer and $V_{OC}$ are reduced. This decreases the efficiency of charge generation and transport reducing $J_{SC}$. As a result, the cell efficiency decreases; however, it can be somewhat increased by doping.

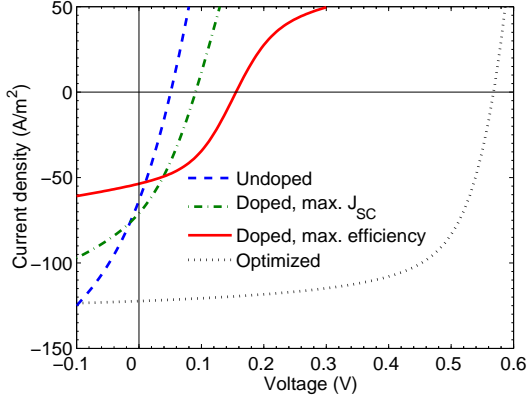

FIG. 12. Calculated $J$-$V$ characteristics for undoped, $p$-doped with $N_p$=0.2x10$^{23}$ m$^{-3}$ (maximum $J_{SC}$) and with $N_p$=1.16x10$^{23}$ m$^{-3}$ (maximum efficiency) bulk heterojunction solar cell with the nonohmic anode contact. For comparison, the $J$-$V$ characteristic for the optimized cell (ohmic contacts) is given.

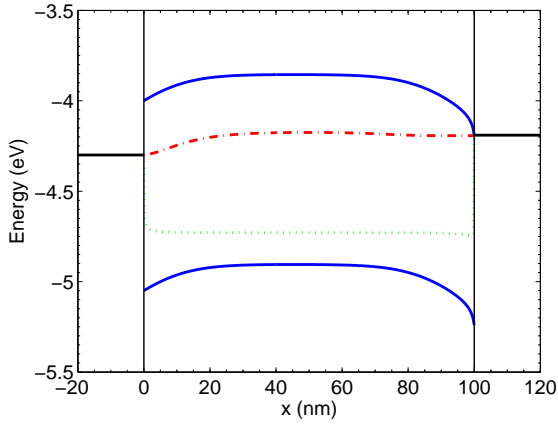

FIG. 13. Energy band diagram at $\Phi_1$=4.3 eV and $p$-dopant concentration $N_p$=1.16x10$^{23}$ m$^{-3}$ in the maximum power point. The lines are explained in Fig. 2.

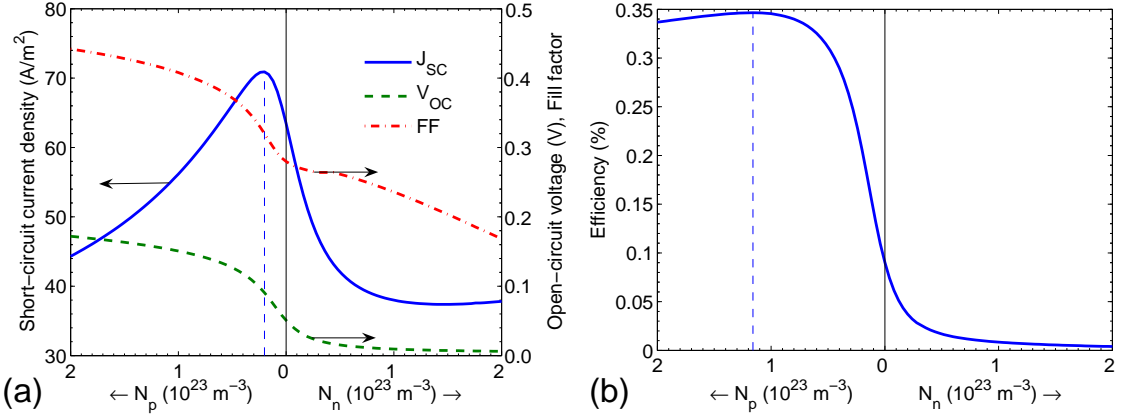

FIG. 14. Short-circuit current density $J_{SC}$, open-circuit voltage $V_{OC}$, fill factor $FF$ (a) and efficiency (b) vs *n*-type and *p*-type dopant concentrations for bulk heterojunction solar cell with nonohmic anode contact. The vertical dashed lines indicate the maximum $J_{SC}$ and efficiency.

With *p* doping, due to the Fermi level difference between the active layer and electrodes the Schottky barriers are formed near the contacts, as follows from Fig. 13. This leads to an increase of the electric field in near-contact regions of the active layer, and, consequently, the efficiency of charge generation and the photocurrent. Figure 14 shows the dependences of $J_{SC}$, $V_{OC}$, $FF$, and the efficiency on *n*- and *p*-dopant concentrations for bulk heterojunction cell with the non-ohmic anode contact. While $V_{OC}$ and $FF$ increase monotonously with p-doping, $J_{SC}$ reaches its maximum at $N_p=0.2 \times 10^{23}$ m$^{-3}$. This behavior of $J_{SC}$ can be explained by thinning the Schottky barriers with doping, thereby decreasing the region of the effective charge generation. As a result, the efficiency is maximal at $N_p=1.16 \times 10^{23}$ m$^{-3}$, and it is ~3.5 times higher than that of the undoped cell.

At *n* doping, the device loses its rectifying ability because both the electrodes are, in fact, electron injecting. Therefore, $V_{OC}$ decreases rapidly to zero as a function of *n*-dopant concentration, thereby nullifying the solar cell efficiency [Fig. 14(b)].

## IV. DOPED BILAYER SOLAR CELLS

The input parameters used in modeling planar heterojunction (bilayer) organic solar cells are almost the same to those used in modeling bulk heterojunction cells (Table I); the differences are given in Table II. These parameters correspond to a bilayer solar cell with P3HT (donor) and PCBM (acceptor) layers. We analyze bilayer cells with strongly ($K=0.1$) and weakly ($K=100$) bound electron-hole pairs generated at the donor-acceptor interface [see Eq. (15)].

TABLE II. Parameters used in modeling bilayer organic solar cells. Index 1 corresponds to P3HT (donor) layer, index 2 corresponds to PCBM (acceptor) layer.

| Parameter | Symbol | Numerical value |
| --- | --- | --- |
| Donor layer thickness | $L_1$ | 50 nm |
| Acceptor layer thickness | $L_2$ | 50 nm |
| Band gap | $E_{g1}$ | 1.85 eV |
|  | $E_{g2}$ | 2.1 eV |
| Electron affinity | $\chi_1$ | 3.2 eV |
|  | $\chi_2$ | 4.0 eV |
| Dielectric constant | $\varepsilon_1$ | 3 |
|  | $\varepsilon_2$ | 4 |
| Surface generation rate of e/h pairs | $G_X$ | $3\times10^{19}$ m$^{-2}$ s$^{-1}$ |
| Dissociation rate/decay rate ratio | $K=k_{diss}(0)/k_f$ | 100; 0.1 |

### A. Doping by majority carriers

Free electrons and holes generated at the donor-acceptor interface move in the acceptor and donor layers, respectively. The majority carriers are electrons in the acceptor layer and holes in the donor layer. Doping by majority carriers here means that the donor layer is *p* doped with concentration $N_{p1}$, and the acceptor layer is *n* doped with concentration $N_{n2}$. The dopant concentrations in each layer are chosen to be equal: $N_{p1}=N_{n2}$.

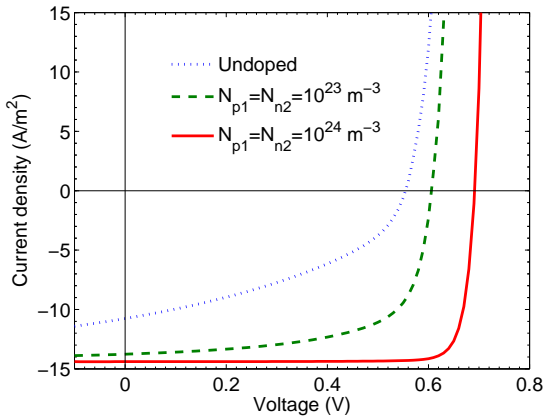

FIG. 15. *J-V* characteristics of the bilayer solar cell at *K*=0.1 and different dopant concentrations.

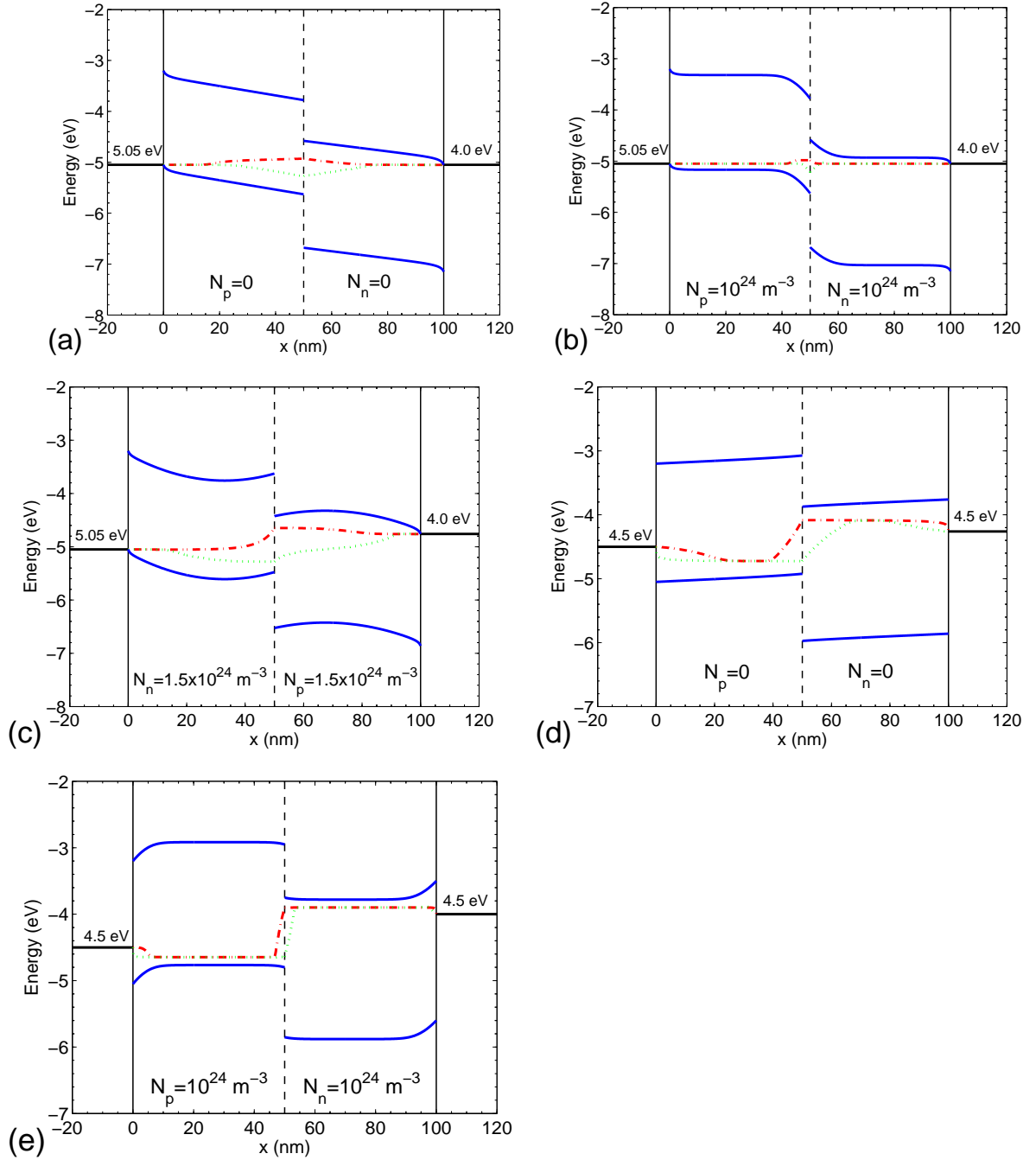

FIG. 16. Band diagrams of bilayer solar cells. The vertical black dashed line denotes the donor-acceptor interface, other lines are explained in Fig. 2. (a) Undoped cell with ohmic contacts; (b) both layers are doped by majority carriers ($10^{24}$ m$^{-3}$), ohmic contacts; (c) both layers are doped by minority carriers (1.5x$10^{23}$ m$^{-3}$), ohmic contacts. Undoped (d) and doped (e) cell with non-ohmic contacts (the electrode work functions $\Phi_1=\Phi_2=4.5$ eV), the doping is as in panel (b). Panels (a) and (b) present the band diagrams at the short-circuit conditions, panels (c), (d) and (e)

show the band diagrams in the maximum power points at voltages 0.29, 0.24 and 0.50 V respectively.

Figure 15 displays the calculated *J-V* characteristics of the bilayer solar cell at different dopant concentrations. The data are plotted for strongly bound electron hole pairs ($K=0.1$) as the doping-induced changes in the *J-V* curves in this case are most pronounced. The case of weakly bound electron-hole pairs ($K=100$) is considered below. Figure 15 shows that doping by majority carriers increases all the solar cell parameters ($J_{SC}$, $V_{OC}$, *FF*, and the efficiency). To explain the effect of doping, consider the cell in short circuit. Figure 16 plots the calculated band diagrams and quasi-Fermi levels for undoped and doped cells. The energy levels in the doped cell are bended near the interface (i.e., at $x=50$ nm) due to exchange of free charges as in *p-n* junction. Therefore the interfacial electric field is much higher in the doped cell. This field grows with increasing the dopant concentrations $N_{p1}$ and $N_{n2}$ leading to an increased dissociation rate of bound electron-hole pairs $k_{diss}(E)$. At a certain dopant concentrations, $k_{diss}(E)$ becomes higher than the decay rate $k_f$; therefore, most bound electron-hole pairs dissociate into free charge carriers, with recombination through the interface suppressed. As a result, $J_{SC}$, $V_{OC}$, and *FF* increase. Figure 17 shows the interface electric field strength at the donor-acceptor interface and $J_{SC}$ versus dopant concentrations. While the interface electric field increases, $J_{SC}$ saturates to its maximum value $eG_X=14.4$ A/m$^2$. At concentrations of $N_{p1}=N_{n2}=2.3 \times 10^{23}$ m$^{-3}$ $J_{SC}$ reaches 99% of the maximum value.

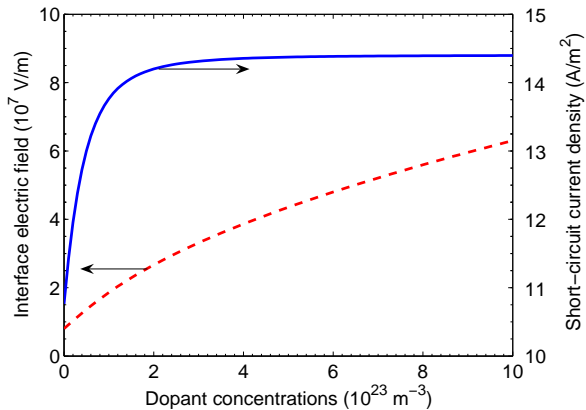

FIG. 17. Electric field strength at the donor-acceptor interface ($x=50$ nm) (dashed line) and $J_{SC}$ (solid line) vs dopant concentrations $N_{p1}=N_{n2}$.

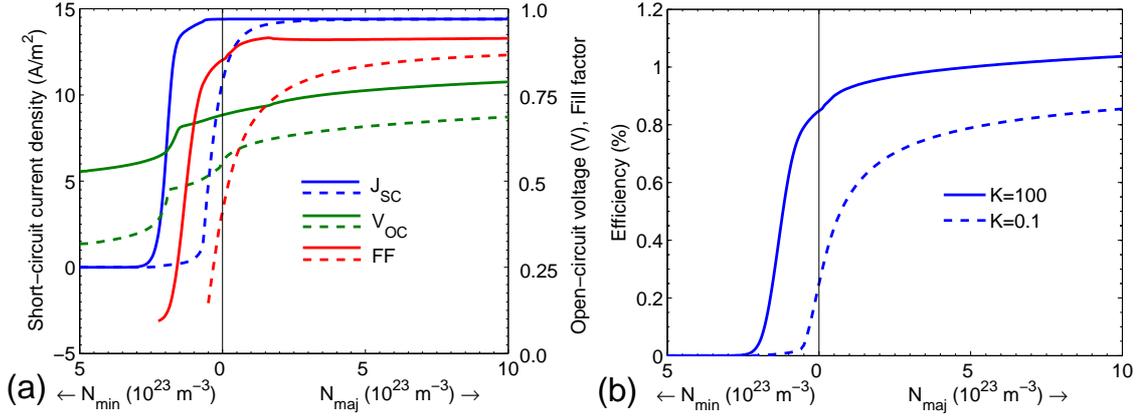

FIG. 18. $J_{SC}$, $V_{OC}$, $FF$ (a) and efficiency (b) for bilayer cells vs. dopant concentrations for majority carrier doping $N_{maj} \equiv N_{p1} = N_{n2}$ and for minority carrier doping $N_{min} \equiv N_{n1} = N_{p2}$. The solid lines correspond to weakly bound electron-hole pairs with $K=100$, the dashed lines correspond to strongly bound electron-hole pairs with $K=0.1$.

Figure 18 presents the dependences of $J_{SC}$, $V_{OC}$, $FF$, and efficiency of bilayer cells on dopant concentrations at $K=0.1$ and 100. As discussed above, for strongly bound electron-hole pairs ($K=0.1$, dashed lines), all the parameters increase with doping by majority carriers. The efficiency increases by 3.4 times at doping mainly due to an increase of $FF$, which enhances by 109%. $J_{SC}$ and $V_{OC}$ grow by 34% and 23%, respectively. For weakly bound electron-hole pairs ($K=100$), the efficiency increase is not so pronounced. Indeed, $J_{SC}$ is already maximal at zero dopant concentration. The slight efficiency growth with doping is mainly due to an increase of $V_{OC}$ and $FF$ by 14% and 7%, correspondingly.

Thus doping by majority carriers improves the performance of bilayer organic solar cells. This is in accordance with the experimental data:[38] it was observed that when the acceptor material was purified, $J_{SC}$, $V_{OC}$, $FF$ of the bilayer organic solar cells were significantly lower than those prepared with as-synthesized materials. The authors suggested that the acceptor material is unintentionally doped, and the purification eliminates doping. Therefore, the interface electric field and the efficiency of free charge generation decrease.[13,38] Note that although our model predicts that doping can enhance the performance of plane heterojunction solar cells, the charge dopants can quench excitons and decrease the charge mobility. These effects can decrease the solar cell performance, but they are out of the scope of the present model. It should be noted that small-molecular materials used for plane heterojunction solar cells are usually purified by gradient sublimation that generally improves the solar cell performance.[39]

However, gradient sublimation does not necessary decrease the doping density of acceptor materials.[38] These data imply that the interrelation between doping, purification, and solar cell performance deserves further studies.

### B. Doping by minority carriers

Doping by minority carriers means that the donor layer is *n* doped with concentration $N_{1n}$, the acceptor layer is *p* doped with concentration $N_{2p}$. As shown in Fig. 18, for minority carrier doping, all the parameters decrease with increasing the dopant concentrations. The photocurrent falls to zero with dopant concentrations, because doping by minority carriers, in contrast with doping by majority carriers, results in decreasing interface electric field, thus reducing the dissociation rate of bound electron-hole pairs.

To illustrate how minority carrier doping can degrade the performance of an optimized bilayer cell, consider the results for weakly bound electron-hole pairs ($K=100$). With doping, *FF* decreases first of all, then $J_{SC}$ and $V_{OC}$ also begin to decrease [Fig. 18(a)].

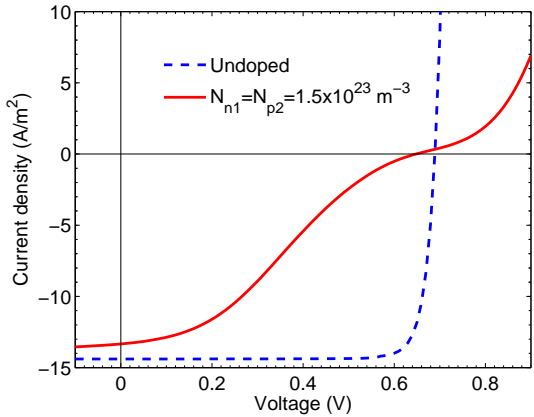

FIG. 19. *J-V* characteristics of the optimized bilayer cell at $K=100$ for undoped (dashed line) and doped (solid line) layers. The layers are doped by minority carriers with concentrations $N_{n1}=N_{p2}=1.5\times10^{23}$ m$^{-3}$.

Figure 19 illustrates the calculated *J-V* characteristics for undoped and doped cells. $J_{SC}$ and $V_{OC}$ change slightly, but *FF* decreases strongly with doping so that the *J-V* curve becomes *S*-shaped. This happens because of the following two reasons. First, in contrast to doping by majority carriers, the interface electric field decreases with doping and even changes its sign. Figure 16(c) shows the band diagram for the doped cell: the slope of the energy levels near the donor-acceptor interface is reversed as compared to the undoped cell [see Fig. 16(a)]. This decreases the dissociation rate of bound electron-hole pairs (the generation rate of free charges) reducing the photocurrent. In addition, the generated free charges while passing through the layers to electrodes undergo bimolecular recombination with charge carriers induced by doping. Figure 20 plots the bimolecular recombination rate $R=\alpha(np-n_0p_0)$ in the bulk of layers. For the undoped cell (dashed line), $R$ exponentially decreases with distance from the interface (vertical black dashed line). If doping by minority carriers is introduced, $R$ (solid line) becomes by many orders of magnitude higher in the bulk. Therefore fewer carriers reach the electrodes and the photocurrent is lowered even further.

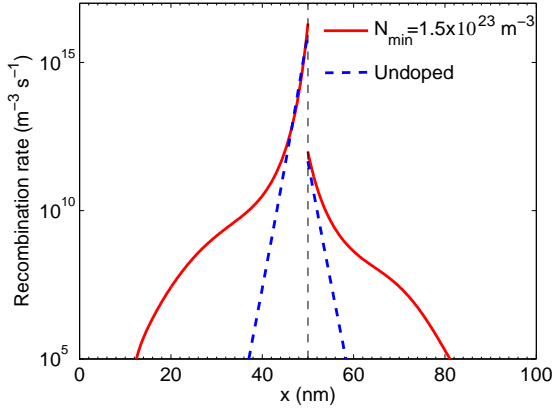

FIG. 20. Rate of bimolecular recombination in the bilayer cell in the maximum power point.

*S*-shaped *J-V* characteristics are occasionally observed in organic solar cells. This was assigned to various reasons: slow charge carrier transfer at the contacts with electrodes,[40,41] the presence of interfacial dipoles, traps, and defects,[42] as well as energy barriers at the donor-acceptor interface.[43] The present study highlights another explanation of the *S*-shaped *J-V* curves – by "inverse doping" of the active layers. Note that the results obtained for bilayer cells can be applicable for bulk heterojunction cells with vertical phase separation.[44]

### C. Nonohmic contacts

To analyze the doping effect on a bilayer solar cell with nonohmic contacts, we use both electrodes with the same work functions ($\Phi_1=\Phi_2=4.5$ eV) so that injection barriers are present at both contacts. We consider the case of weakly bound electron-hole pairs ($K=100$), with the other parameters being the same as in the previous sections. Figures 16 (d) and 16 (e) present the energy band diagrams at the maximum power point for the nonohmic bilayer cell with undoped layers and doped ones by majority carriers with dopant concentrations of $10^{24}$ m$^{-3}$, respectively. In the undoped cell, the electric field is constant through the active layer, and its direction is opposite to the photocurrent at positive voltages. This field suppresses dissociation of the bound electron-hole pairs at the interface and prevents charge transport to the electrodes. This is in accordance with *J-V* curves calculated for the bilayer cell with nonohmic contacts at different dopant levels and shown in Figure 21. Indeed, at voltages higher than 0.2 V, the non-ohmic contacts strongly reduce the photocurrent and, hence, *FF* of the undoped cell is very small. Note that, at the low voltages, the photocurrent is insensitive to the non-ohmic contacts as a result of efficient diffusive transport of the dissociated electron-hole pairs to the electrodes. When the doping is introduced [see Fig. 16 (e)], a positive electric field appears at the interface, and the opposite electric field is screened in the bulk of the layers. As a result, *FF* considerably increases with dopant concentrations (see Fig. 21). Therefore, the doping strongly enhances the performance of the nonohmic bilayer cell, and it brings its efficiency (0.69% at doping level $10^{24}$ m$^{-3}$) to about that of the optimized cell (0.84%). As a result of doping, $V_{OC}$ and *FF* increase from 0.66 V and 31.2% to 0.80 V and 59.7% so that the efficiency increases more than twice (from 0.29% to 0.69%). Note that the contacts in the doped cell stay non-ohmic according to our model.

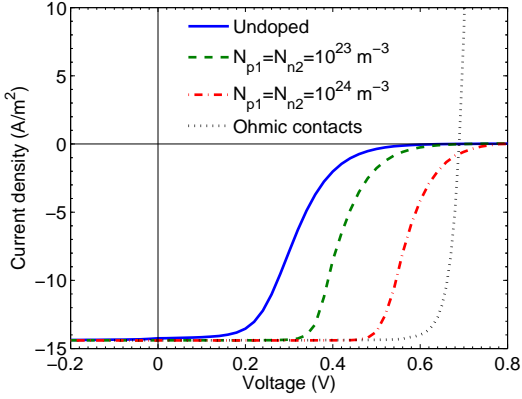

FIG. 21. *J-V* characteristics for bilayer solar cells with non-ohmic contacts ($\Phi_1=\Phi_2=4.5$ eV) for undoped (solid line), doped by majority carriers with concentrations $10^{23}$ m$^{-3}$ and $10^{24}$ m$^{-3}$ (dashed and dot-dashed lines, respectively) layers. The dotted line is the *J-V* curve for the undoped optimized cell with ohmic contacts. The data are calculated for weakly bound electron-hole pairs (*K*=100).

The effect of non-ohmic contact could also explain the *S*-type *J-V* curves from Ref. [38] discussed in Sec. IV A. The authors used an Ag top electrode on the acceptor layer so that a barrier (i.e., non-ohmic contact) could prevent efficient electron injection in the acceptor layer and result in *S*-type *J-V* curves with the undoped (purified) acceptor materials. According to our model, doping (using the unpurified or air-exposed materials as suggested in Ref. [38]) increases the interface field that leads to increased $V_{oc}$, *FF* and, hence, performance.

.

## V. CONCLUSION

We have extended the known numerical models of bulk and planar heterojunction organic solar cells by introducing doping of the active layer(s) and studied their performance. Solar cells based on the material pair P3HT-PCBM were modeled.

We have found that for the optimized bulk heterojunction solar cell doping degrades its performance: $J_{SC}$ and $FF$ decrease with either *n*- or *p*-doping because of reducing the electric field in the active layer; $V_{OC}$ changes slightly. For non-optimized bulk heterojunction cells, namely for low or unbalanced charge carrier mobilities, or non-ohmic contacts, the efficiency can be increased by doping. For low charge mobilities and non-ohmic contacts, due to the Schottky barrier formation, doping increases the electric field enhancing the dissociation rate of bound electron-hole pairs. At unbalanced mobilities, doping compensates the space charge caused by slower carriers increasing the electric field and photocurrent.

For bilayer organic solar cells doping by majority carriers (i.e., both *n*-doping the acceptor layer and *p*-doping the donor one) enhances $J_{SC}$, $V_{OC}$, $FF$ and, hence, the performance because of increasing the interfacial electric field. Inversely, i.e., with doping by minority carriers, $FF$ significantly decreases that can result in an *S*-shaped *J-V* curve. Moreover, the performance of bilayer solar cells with non-ohmic contacts can be strongly enhanced by proper doping.

Doping can strongly influence the performance of organic solar cells and should be taken into account in their optimization. The developed numerical device model could serve a useful tool for this purpose. To increase the range of dopant concentrations at which the model is valid, the effect of doping on exciton quenching and charge mobility should be taken into account.

## VI. ACKNOWLEDGMENT

This work was supported by the Ministry of Education and Science of the Russian Federation (contracts Nos. 16.740.11.0064, 16.740.11.0249, 02.740.11.5155, and 11.519.11.6020).

**References**

1. G. G. Malliaras, J. R. Salem, P. J. Brock, and J. C. Scott, J. Appl. Phys. **84**, 1583 (1998).
2. P. Schilinsky, C. Waldauf, J. Hauch, and C. J. Brabec, J. Appl. Phys. **95**, 2816 (2004).
3. L. J. A. Koster, V. D. Mihailetchi, R. Ramaker, and P. W. M. Blom, Appl. Phys. Lett. **86**, 123509 (2005).
4. J. A. Barker, C. M. Ramsdale, and N. C. Greenham, Phys. Rev. B **67**, 075205 (2003).

[5] G. Yu, K. Pakbaz, and A. J. Heeger, Appl. Phys. Lett. **64**, 3422 (1994).

[6] S. E. Shaheen, C. J. Brabec, N. S. Sariciftci, F. Padinger, T. Fromherz, and J. C. Hummelen, Appl. Phys. Lett. **78**, 841 (2001).

[7] J. K. J. van Duren, X. Yang, J. Loos, C. W. T. Bulle-Lieuwma, A. B. Sieval, J. C. Hummelen, and R. A. J. Janssen, Adv. Funct. Mater. **14**, 425 (2004).

[8] C. J. Brabec, N. S. Sariciftci, and J. C. Hummelen, Adv. Funct. Mater. **11**, 15 (2001).

[9] L. J. A. Koster, E. C. P. Smits, V. D. Mihailetchi, and P. W. M. Blom, Phys. Rev. B **72**, 085205 (2005).

[10] C. Deibel, A. Wagenpfahl, and V. Dyakonov, Phys. Status Solidi (RRL) **2**, 175 (2008).

[11] A. Wagenpfahl, C. Deibel, and V. Dyakonov, IEEE J. Sel. Top. Quantum Electron. **99**, 1 (2010).

[12] B. A. Gregg, J. Phys. Chem. C **113**, 5899 (2009).

[13] B. A. Gregg, Soft Matter **5**, 2985 (2009).

[14] P. Peumans, A. Yakimov, and S. R. Forrest, J. Appl. Phys. **93**, 3693 (2003).

[15] B. A. Gregg, S.-G. Chen, and R. A. Cormier, Chem. Mater. **16**, 4586 (2004).

[16] B. A. Gregg and R. A. Cormier, J. Am. Chem. Soc. **123**, 7959 (2001).

[17] G. Dicker, M. P. de Haas, J. M. Warman, D. M. de Leeuw, and L. D. A. Siebbeles, J. Phys. Chem. B **108**, 17818 (2004).

[18] B. A. Gregg, S. E. Gledhill, and B. Scott, J. Appl. Phys. **99**, 116104 (2006).

[19] A. J. Mozer, N. S. Sariciftci, A. Pivrikas, R. Österbacka, G. Juška, L. Brassat, and H. Bässler, Phys. Rev. B **71**, 035214 (2005).

[20] D. Wang, M. O. Reese, N. Kopidakis, and B. A. Gregg, Chem. Mater. **20**, 6307 (2008).

[21] T. Tromholt, E.A. Katz, B. Hirsch, A. Vossier, and F. C. Krebs, Appl. Phys. Lett. **96**, 073501 (2010).

[22] J. G. Xue, S. Uchida, B. P. Rand, S. R. Forrest, Appl. Phys. Lett. **84**, 3013 (2004).

[23] P. Langevin, Ann. Chim. Phys. **28**, 433 (1903).

[24] C. Deibel, A. Wagenpfahl, and V. Dyakonov, Phys. Rev. B **80**, 075203 (2009).

[25] S. Selberherr, *Analysis and Simulation of Semiconductor Devices* (Springer-Verlag, Wien, New York, 1984).

[26] C. L. Braun, J. Chem. Phys. **80**, 4157 (1984).

[27] L. Onsager, Phys. Rev. **54**, 554 (1938).

[28] T. E. Goliber and J. H. Perlstein, J. Chem. Phys. **80**, 4162 (1984).

[29] V.D. Mihailetchi, L. J. A. Koster, J. C. Hummelen, and P.W.M. Blom, Phys. Rev. Lett. **93**, 216601 (2004).

[30] H. K. Gummel, IEEE Trans. Electron Devices **11**, 455 (1964).

[31] D. L. Scharfetter and H.K. Gummel, IEEE Trans. Electron Devices **16**, 64 (1969).

[32] A. K. Jonscher, Thin Solid Films **1**, 23 (1967).

[33] P. W. M. Blom, V. D. Mihailetchi, L. J. A. Koster, and D. E. Markov, Adv. Mater. **19**, 1551 (2007).

[34] A. M. Goodman and A. Rose, J. Appl. Phys. **42**, 2823 (1971).

[35] V. D. Mihailetchi, P. W. M. Blom, J. C. Hummelen, and M. T. Rispens, J. Appl. Phys. **94**, 6849 (2003).

[36] H. Frohne, S. E. Shaheen, C. J. Brabec, D. C. Müller, N. S. Sariciftci, and K. Meerholz, ChemPhysChem **3**, 795 (2002).

[37] J. S. Kim, M. Granstörm, and R. H. Friend, J. Appl. Phys. **84**, 6859 (1998).

[38] A. Liu, S. Zhao, S.-B. Rim, J. Wu, M. Könemann, P. Erk, and P. Peumans, Adv. Mater. **20**, 1065 (2008).

[39] R. F. Salzman, J. Xue, B. P. Rand, A. Alexander, M. E. Thompson, S. R. Forrest, Org. Electron. **6**, 242 (2005).

[40] M. Glatthaar, M. K. Riede, N. Keegan, K. Sylvester-Hvid, B. Zimmermann, M. Niggemann, A. Hinsch, and A. Gombert, Sol. Energy Mater. Sol. Cells **91**, 390 (2007).

[41] A. Wagenpfahl, D. Rauh, M. Binder, C. Deibel, and V. Dyakonov, Phys. Rev. B **82**, 115306 (2010).

[42] A. Kumar, S. Sista, and Y. Yang, J. Appl. Phys. **105**, 094512 (2009).

[43] C. Uhrich, D. Wynands, S. Olthof, M. K. Riede, K. Leo, S. Sonntag, B. Maennig, and M. Pfeiffer, J. Appl. Phys. **104**, 043107 (2008).

[44] L. M. Chen, Z. Xu, Z. R. Hong, and Y. Yang, J. Mater. Chem. **20**, 2575 (2010).